\documentstyle[prc,aps,preprint]{revtex}
\begin{document}
\draft
\date{\today}
\title{A Theoretical Study of $\rho^{0}$-Photoproduction on Nucleons
near Threshold}
\author{H. Babacan, T. Babacan, A. Gokalp, O. Yilmaz}
\address{Physics Department, Middle East Technical University,
06531 Ankara, Turkey}
\maketitle
\begin{abstract}

We investigate the possibility that the process of
$\rho^{0}$-meson photoproduction on proton, $\gamma+p\rightarrow
p+\rho^{0}$, in the near threshold region $E_{\gamma}< 2$ GeV, can
be considered in the framework of model with $\pi$-, $\sigma$- and
N-exchanges. This suggestion is based on a study of the
t-dependence of differential cross section, $d\sigma(\gamma p
\rightarrow p \rho^{0})/dt$, which has been measured by SAPHIR
Collaboration. We find that the suggested model provides a good
description of the experimental data with new values of $\rho
NN$-coupling constants  in the region of the time-like
$\rho^{0}$-meson momentum. Our results suggest that such model can
be considered as a suitable nonresonant background mechanism for
the future discussion of possible role of nucleon resonance
contributions. Our predictions for $\rho^{0}$-meson
photoproduction on neutron target and for beam asymmetry on both
proton and neutron targets are presented.
\end{abstract}

\thispagestyle{empty} ~~~~\\ \pacs{PACS numbers:
13.60.Le;13.60.-r;13.88.+e;24.70.+s;25.20.Lj}
%\narrowtext
\newpage

\section{Introduction}

The photoproduction of $\rho$- and $\omega$-mesons on nucleons,
$\gamma+N\rightarrow N+V$, near threshold $E_{\gamma}< 2$ GeV, is
considered typically as  a possible way for the study of the
physics of nucleon resonances $N^{*}$ in the interesting dense
region of its masses, $M_{N^{*}} > M+m_{v} = 1 .7$ GeV,
where $M(m_{v})$
is the nucleon (vector meson) mass. Especially, such experiments
could be interesting for the search and subsequent study of the
so-called missing resonances \cite{R1,R2}. Typical opinion here is
that due to possible large width of the decay $N^{*}\rightarrow
N+V(\rho, \omega)$, the reactions of vector meson
photoproduction on nucleons will be sensitive to these
resonances. Therefore,
the future intensive flux of new data from JLAB will be effective
for the solution of this problem. Multipole analysis of experimental
data about different observables in processes $\gamma+N\rightarrow
N+V$ can be realized only in the case of an appropriate and a realistic
model for the nonresonant mechanisms for $\gamma+N\rightarrow
N+V$. This is especially important for the photoproduction of
neutral vector mesons, where $N^{*}$ contributions do not seem
as the main ones \cite{R2,R3,R4}.

In the literature \cite{R1,R2,R3,R4,R5,R6,R7,R8,R9,R10} the
following nonresonant
mechanisms are discussed: the pseudoscalar ($\pi$,$\eta$) and
scalar ($\sigma$) exchanges in t-channel, one nucleon
exchanges in s- and u-channels, and Pomeron exchange. This
introduces a large enough set of unknown parameters,
characterizing different contributions, such as the coupling
constants, their relative phases, and the cut-off parameters of
numerous phenomenological form factors, as well. In principle
different combinations of these ingredients are
presented in the literature. For example, in Ref.\cite{R5} the
model for $\gamma+N\rightarrow N+\rho(\omega)$ contains the
following two contributions: ($\pi+\sigma$)-exchanges in t-channel;
with specific form factors in electromagnetic and strong vertices
of pole diagrams. The same nonresonant background, i.e.
$(\pi+\sigma)$-exchanges, is also considered in Ref.\cite{R2}, with the
same coupling constants but with different form factors. The
corresponding model in Ref.\cite{R3} contains three ingredients:
$(\pi+\sigma)$-exchanges in t-channel, (s+u)-channel one-nucleon
contribution
and Pomeron exchange.

Our aim here is to suggest a simple enough model for the process
$\gamma+N\rightarrow N+\rho^{0}$ in the near threshold region
which will describe relatively well the existing experimental data
\cite{R11} about differential cross sections for
$\gamma+p\rightarrow p+\rho^{0}$ and will produce nontrivial
polarization phenomena, more rich, for example, than in the case
of $(\pi+\sigma)$-exchange. For such exchanges almost all
polarization phenomena are trivial and can be predicted without
knowledge of exact values of the coupling constants and
phenomenological form factors. For example, the beam asymmetry
$\Sigma$ induced by the linear polarization of the photon beam, and
all possible T-odd polarization observables such as, for example,
target asymmetry or polarization of final proton produced in
collisions of unpolarized particles will be zero identically for
any kinematical conditions of the considered reaction.
Analogously, it is possible to predict that $\rho_{11} = 1$, and
all other elements of the $\rho$-meson density matrix must be
zero. Let us note also that $(\pi+\sigma)$-model will not produce any
difference in cross section on proton and neutron targets due to
the absence of $\sigma$- and $\pi$-interference.

But the suggested  model in this work   will be more rich and more
flexible,allowing to predict nontrivial polarization phenomena.
Such model
($\pi+\sigma+N$) will be suitable enough as starting point for the
discussion of possible contribution of nucleon resonances. And
presence of different interference contributions such as
$\sigma\bigotimes N$  and $\pi\bigotimes N$   in the
differential cross section even with unpolarized particles will be
important for establishing relative signs of coupling constants.
That will be crucial for more complete information about these
constants, which is necesssary for prediction of polarization
phenomena.

\section{Exchange Mechanisms and Amplitudes}

For t-channel, we consider the pseudoscalar ($\pi$), and scalar ($\sigma$)
exchanges, shown in Fig.1a. The pseudoscalar exchange
amplitudes
can be obtained from the Lagrangian,
\begin{eqnarray}
{\cal
L}_{\pi}=\frac{e}{m_{\rho}}g_{\rho\pi\gamma}\epsilon^{\mu\nu\alpha\beta}
\partial_{\mu}V_{\nu}\partial_{\alpha}A_{\beta}\pi-ig_{\pi NN}\bar{N}\gamma_5
N,
\end{eqnarray}
where  and $A_{\mu}$($V_{\mu}$) is the photon (vector meson) field.
Then, one-pion exchange amplitude takes the form
\begin{equation}
 {\cal M}_{t}
 =e~\frac{g_{\rho\pi\gamma}}{m_{\rho}}~\frac{g_{\pi NN}}{t-m_{\pi}^{2}}
 F_{\pi NN}(t)~F_{\rho\pi\gamma}(t)~
 (\overline{u}(p_{2})~\gamma_{5}~u(p_{1}))
 ~(\epsilon^{\mu\nu\alpha\beta}
 ~\varepsilon_{\mu}~k_{\nu}~U_{\alpha}~q_{\beta}),
 \nonumber \\ \nonumber \\
\end{equation}
where $t=(k-q)^{2}$, $m_{\rho}$ is the mass of $\rho^{0}$-meson,
$~\varepsilon_{\mu}(U_{\mu})$ is the polarization four vector of
photon(vector meson). Notation of particle four momenta is
presented in Fig. 1. We shall use the coupling constants as
$g_{\rho\pi\gamma}=0.54$ and $g_{\pi NN}^2/4\pi=14.0$. The coupling
constant $g_{\rho\pi\gamma}$ is obtained from the experimental
partial width of $\rho^{0}$ radiative decay
$\rho^{0}\rightarrow\pi^{0}+\gamma$ \cite{R12}. The form factors used in
our calculations are
\begin{eqnarray}
F_{\pi
NN}(t)=\frac{\Lambda_{\pi}^2-m_{\pi}^2}{\Lambda_{\pi}^2-t}~~,
F_{\rho\pi\gamma}(t)=\frac{\Lambda_{\rho\pi\gamma}^2-m_{\pi}^2}{\Lambda_{\rho\pi\gamma}^2-t},
\end{eqnarray}
where $\Lambda_{\pi}=0.7$~GeV and $\Lambda_{\rho\pi\gamma}=0.77$
~GeV \cite{R5}.

The scalar ($\sigma$) exchange amplitude can be obtained from the
Lagrangian,
\begin{eqnarray}
 {\cal L}_{\sigma}=\frac{e}{m_{\rho}}g_{\rho\sigma\gamma}
 (\partial^{\alpha}V^{\beta}\partial_{\alpha}A_{\beta}-\partial^{\alpha}V^
 {\beta}\partial_{\beta}A_{\alpha})\sigma
 +g_{\sigma NN}\bar{N}N\sigma.
\end{eqnarray}
The above Lagrangian leads to  the following expression for scalar
exchange amplitude:
\begin{eqnarray}
 {\cal M}_{\sigma}& = & e \frac{g_{\rho\gamma\sigma}}{m_{\rho}}
 \frac{g_{\sigma NN}}{t-m_{\sigma}^{2}}F_{\sigma NN}(t)
 F_{\rho\sigma\gamma}(t)(\overline{u}(p_{2})\gamma_{5}u(p_{1}))
 (\varepsilon\cdot U~k\cdot q-\varepsilon\cdot q~U\cdot k),\nonumber \\
\end{eqnarray}
where $g_{\rho\gamma\sigma}$ and $g_{\sigma N N}$ are the coupling
constants for the vertices $\rho\gamma\sigma$ and $\sigma N N$.
Following Ref.\cite{R5}, they are taken as $g_{\sigma
NN}^2/4\pi=8.0$, and $g_{\rho\gamma\sigma}=2.7$. The form factors for this
exchange mechanism are given by
\begin{eqnarray}
F_{\sigma
NN}(t)=\frac{\Lambda_{\sigma}^2-m_{\sigma}^2}{\Lambda_{\sigma}^2-t}~~,
F_{\rho\sigma\gamma}(t)=\frac{\Lambda_{\rho\sigma\gamma}^2-m_{\sigma}^2}
{\Lambda_{\rho\sigma\gamma}^2-t},
\end{eqnarray}
where $\Lambda_{\sigma}=1.0$~GeV and
$\Lambda_{\rho\sigma\gamma}=0.9$~GeV \cite{R5}.

The Lagrangian for VNN and $\gamma NN$ interactions can be written
in the following way:
\begin{eqnarray}
{\cal L}_{N}=\bar{N}(g_{\rho NN}^{V}
\hat{V}-\frac{g_{\rho NN }^{T}}{2
M}\sigma_{\mu\nu}\partial^{\nu}V^{\mu})N
        +e\bar{N}(Q_N\hat{A}-\frac{\kappa_N}{2 M
        }\sigma_{\mu\nu}\partial^{\nu}A^{\mu})N,
\end{eqnarray}
where we use the notation $\hat{a}=a_{\mu}\gamma^{\mu}$, and
$\sigma_{\mu}=(\gamma_{\mu}\gamma_{\nu}-\gamma_{\nu}\gamma_{\mu})/2$. The
s- and u-channel amplitudes can then be obtained by using the
above Lagrangian as
\begin{eqnarray}
{\cal M}_{s}=e \frac{g_{\rho NN }^{V}}{s-M^{2}}
\overline{u}(p_{2})(\hat{U}+ \frac{\kappa_{\rho}}{2 M }
\hat{U}~\hat{q})(\hat{p}+M) (Q_{N} \hat{\varepsilon}-
\frac{\kappa_{N}}{2 M} \hat{\varepsilon}~\hat{k})u(p_{1}),\nonumber \\
\end{eqnarray}
\begin{eqnarray}
{\cal M}_{u}=e \frac{g_{\rho NN }^{V}}{u-M^{2}}
\overline{u}(p_{2})(Q_{N} \hat{\varepsilon}- \frac{\kappa_{N}}{2M}
\hat{\varepsilon}~\hat{k}) (\hat{f}+M) (\hat{U}+
\frac{\kappa_{\rho}}{2M}~\hat{U}~\hat{q}) u(p_{1}),\nonumber\\
\end{eqnarray}
where $s=(k+p_{1})^{2}$, $u=(k-p_{2})^{2}$, $f=p_{1}-q$,
$p=p_{2}+q$, $g_{\rho NN}^{V}$ and $g_{\rho NN}^{T}$ are the vector
and tensor coupling constants for $\rho NN$-vertex, $Q_{N}=1(0)$
is the electric charge for proton(neutron),
$\kappa_{N}=1.79(-1.91)$ is the anomalous magnetic moment of
proton(neutron),and $\kappa_{\rho}$ is defined as
$\kappa_{\rho}=g_{\rho NN}^{T}/g_{\rho NN}^{V}$. The values of the
coupling constants $g_{\rho NN}^{V}$ and $g_{\rho NN}^{T}$ will
be given in the next section.

Let us note that the suggested model for the matrix element of the
process
$\gamma +N\rightarrow N+V$ namely ${\cal M}$= ${\cal M}_{\pi}+
{\cal M}_{\sigma}+ {\cal M}_{s}+ {\cal M}_{u}$ satisfies the gauge
invariance of hadron electromagnetic interaction at any values of the
coupling constants and form factors in the whole region of
kinematical variables s and t. Futhermore, we like to mention  that we do
not introduce any form factor
in ${\cal M}_{s}$
and ${\cal M}_{u}$. Problem here is that in general the
form factors for ${\cal M}_{s}$ and ${\cal M}_{u}$ must be
different, having s- or u-dependences, respectively. But
this  "natural" form factors will destroy the gauge invariance of
s+u contribution to the total matrix element for $\gamma
+p\rightarrow p+V$. In principle, it is possible to introduce some
common phenomenological form factor in front of ${\cal
M}_{s}+{\cal M}_{u}$ \cite{R13} with s- and u- dependences
simultaneously as F(s,u). Such a "form factor" F(s,u), depending
on two variables seems more like  as some amplitude, but not as form
factor which typically depends on one variable. So this dependence
differs from the case of t-channel where the corresponding form
factors are the functions of t-variable only.

Another point which must be stressed here concerns the values of
the coupling constants, $g_{VNN}^{V}$ and $g_{VNN}^{T}$ for the
vertex $VNN$. Typical way in the literature is to use for these
constants information from NN-interaction \cite{R14,R15} or
pion photoproduction processes, $\gamma +N\rightarrow N+\pi$
\cite{R16}, where vector meson exchange plays some role. But the
case we consider, $\gamma +N\rightarrow N+V$, on one hand,
and $N+N\rightarrow N+N$, for example, on the other hand, are
controlled by the VNN-constants in the different regimes of vector
meson momentum : space-like in the case of NN-interaction or
$\gamma +N\rightarrow N+\pi$ and time-like
for $\gamma +N\rightarrow N+V$. Therefore to
connect $N+N\rightarrow N+N$ and $\gamma +N\rightarrow N+V$, a
long extrapolation in momentum transfer must be done. So
VNN-coupling constants for these cases could be essentially
different. And another important difference in VNN-constants from
different processes, which must be mentioned here, concerns the high
virtuality of one of the nucleons for the VNN-vertex in the case of
processes $\gamma +N\rightarrow N+V$.

These comments could be considered as some justification of our
strategy in the consideration of these coupling constants: namely,
we shall consider these constants as free parameters, whose values
must be adjusted by some fit to the existing experimental data
about differential cross sections  for process
$\gamma+p\rightarrow p+\rho^{0}$ in the near threshold region.

Moreover, in our consideration here we will neglect the Pomeron
contribution to the matrix element for $\gamma +N\rightarrow N+V$
in the near threshold region, $E_{\gamma}< 2$ GeV. It is possible
to justify such approach by observation that the Pomeron, being as
an effective high energy phenomenological phenomenon, does not
seem as the adequate mechanism in the near threshold region. For
example, in another processes, where the Pomeron exchange is
allowed definitely, such as the elastic $\pi+N$, $K+N$ or
N+N-scattering \cite{R17}, its contribution is considered
typically at higher values of invariant variable s in
comparison with the near threshold values of s for $\gamma
+N\rightarrow N+V$. For example, $W_{th}(NN\rightarrow NN)=2M >
W_{th}(\gamma N \rightarrow N V)=M+m_{v}$, where $s=W^{2}$, but it is
evident
that
in $N+N\rightarrow N+N$ the Pomeron exchange is taken into account
at more higher W-values. Therefore, it is difficult to find some
specific theoretical reasons to justify the Pomeron exchange in
the threshold region for processes $\gamma+N\rightarrow N+V$. This
region can be considered as the transition regime from the
contributions of s-channel nucleon resonances to the Regge approach
in accordance with the duality hypothesis \cite{R17,R18,R19}. Namely,
in this region
there is the delicate problem of double counting if both above
mentioned contributions are considered simultaneously. Therefore,
to avoid this problem we will not consider the Pomeron
contribution in the near threshold region $E_{\gamma}\leq 2$ GeV,
for $\gamma+p\rightarrow p+\rho(\omega)$.

We do not consider in our work the nucleon resonances as well.
Diffractive-like behaviour of the differential cross sections for
$\gamma+p\rightarrow p+V$ processes even very near to the
threshold can be considered as some indication that this mechanism
can not be main one here. For example, the analysis in the quark model
of the contribution of large number of nucleon resonances
demonstrated that they cannot reproduce such diffractive
t-dependence \cite{R2}.

Let us note that the contribution of nucleon resonance $N^{*}$ with the
definite value of spin J and parity P, $J^{P}$, to the amplitude
of $\gamma+N\rightarrow N+V$ process is complicated generally, being
characterized by six independent constants or partial amplitudes,
for $J\geq 3/2$. These amplitudes correspond to two possible
initial $(\gamma+N)$-states with the electric and magnetic
multipolarities of real photons in the chain of the transitions:
$\gamma+N\rightarrow N^{*}(J^{P})\rightarrow N+V$ and three
different final $(V+N)$-states with definite combinations of total
VN spin, $S_{f}=1/2$ and 3/2, and the orbital angular momentum of
the vector meson. Even for $J^{P}=(1/2)^{\pm}$ there are two independent
transitions, i.e. the situation with $N^{*}$ in processes
$\gamma+N\rightarrow N+V$ is more complicated in comparison with
$\gamma+N\rightarrow N+\pi$ or $\gamma+N\rightarrow N+\eta$, where
the $N^{*}$ contribution with some $J^{P}$ is characterized by two
multipole amplitudes only.

In the case of the Breit-Wigner parametrization of the
$N^{*}$ contribution to the matrix element for
$\gamma+N\rightarrow N+V$ process, each such contribution is
characterized by five constants: two electromagnetic ones,
magnetic and electric, and three strong constants for the decay
$N^{*}\rightarrow N+V$. In principle, it is possible to use
information about the electromagnetic vertex, $N^{*}\rightarrow
N+\gamma$, from the multipole analysis of processes
$\gamma+N\rightarrow N+\pi(\eta)$. So, the processes
$\gamma+N\rightarrow N+V$ will be used for the study of the spin
structure of strong vertices: $N^{*}\rightarrow N+V$. But it is
not the case for the missing resonances, with the small
$N^{*}\rightarrow N+\pi$ branching ratio, i.e. with a weak signal
in $\gamma+N\rightarrow N+\pi$, with unknown electromagnetic
constants.

It is evident that the successful solution of the missing
resonance problem, using the processes $\gamma+N\rightarrow
N+\rho(\omega)$, needs a large amount of polarization data with
polarized beam, polarized target, with measurements of
polarization properties of produced vector mesons. Only in that
case the corresponding multipole analysis can be done more or less
uniquely.

We like to note that the suggested model here produces real amplitudes
and as a result all possible T-odd polarization observables must
be identically zero, independently on the relative role of the
considered mechanisms. But $N^{*}$-contribution in s-channel
will change the situation qualitatively, introducing a new essential
property, namely complexity of amplitudes with a rich T-odd
polarization phenomena. Therefore these observables will be
especially sensitive to possible $N^{*}$-contribution to the
matrix element for $\gamma+p\rightarrow p+\rho^{0}$. Even not so
intensive $N^{*}$-contribution, through its interference with
large $\sigma$-contribution, can produce a detectable signal in
target asymmetry, for example. But before that the following problem
must be solved: are there some other sources of amplitude
complexity in the near threshold region for $\gamma+p\rightarrow
p+\rho^{0}$? Evidently the complex Pomeron exchange through its
specific signature can not be considered as a good mechanism near
threshold. Of course, it is necessary to keep in mind final
VN-interaction, which will modify the real $\pi$ and $\sigma$
contributions. This way we will not meet the delicate
problem of double counting in the case of additional
$N^{*}$-contributions.

In any case, the problem of missing resonances in $\gamma+N
\rightarrow N+V^{0}$, being as very interesting, will introduce
necessity of solution of some serious problems. And one such
problem is the choice of the adequate model for the nonresonant
mechanism  in $\gamma+N \rightarrow N+V^{0}$ $(\rho^{0},\omega)$,
where namely this nonresonant background is
the main mechanism in the near threshold region. Intensive
study of polarization phenomena in $\gamma+p \rightarrow
p+V^{0}$ will be very important to successful solution of this
and another related problems.

And we must repeat here once more, that the finding an adequate model
for the nonresonant contributions to the matrix element for
$\gamma+p\rightarrow p+\rho(\omega)$ in the near threshold region
is an important task, especially taking into account the numerous
number of possible $N^{*}$ with many fitting free parameters.
Interference of different mechanisms must be intensive enough, and
that will introduce additional problem of relative phases of
different contributions.

\section{Results and Discussion}

Even such relatively simple model contains large number of unknown
parameters: $\Lambda_{\pi}$, $\Lambda_{\sigma}$,
$\Lambda_{\rho\sigma\gamma}$, $\Lambda_{\rho\pi\gamma}$,
$g_{\rho\sigma\gamma}$, $g_{\sigma NN}$, $g_{\rho NN}^{V}$,
$g_{\rho NN}^{T}$. The $\sigma$-meson mass in principle can also
be considered as a parameter, being limited in the wide interval:
$400\leq m_{\sigma}\leq 1200$ MeV \cite{R12}. Only two coupling constants,
namely $g_{\pi NN}$ and $g_{\rho\pi\gamma}$ can be considered as
well known. Evidently any attempt to find all these
parameters by fitting the limited set of experimental data
concerning only differential cross section for process $\gamma
+p\rightarrow p+\rho^{0}$, with unpolarized particles, cannot be
successful. Therefore we follow the literature tradition
\cite{R5}, and we will fix all four cut-off parameters, $\Lambda_{i}$,
choosing their values as it was indicated in Sec. 2.

After this we have only three parameters, $g_{\rho NN}^{V}$,
$g_{\rho NN}^{T}$, and $g_{\rho\sigma\gamma}$, assuming $g_{\sigma
NN}^{2}/4\pi=8.0$, as it follows from NN-interaction \cite{R14}. We
fit our model with all the $d\sigma/dt$ data from SAPHIR Collaboration
at three energy intervals: $1.45<E_{\gamma}<1.64$ GeV,
$1.64<E_{\gamma}<1.82$ GeV, $1.82<E_{\gamma}<2.03$ GeV. Note that
we choose the "standard value" for $\sigma$-mass: $m_{\sigma}=500$
MeV, evidently another value of $m_{\sigma}$ from allowed
wide interval can change the results of fitting.

We choose specially only the above three mentioned from existing six
energy intervals for $d\sigma/dt$, measured by SAPHIR Collaboration
\cite{R11}, to have
the possibility to predict $d\sigma/dt$ for another three energy
intervals, and to test in some sense the validity of the
suggested model.

The best results for the fitted coupling constants are the
following two sets of coupling constants: $g_{\rho NN}^{V}=0.4$, $g_{\rho
NN}^{T}=1.0$ for the
fixed value of coupling constant $g_{\rho\sigma\gamma}$, namely ,
$g_{\rho\sigma\gamma}=2.7$, following Ref. \cite{R5} and $g_{\rho
NN}^{V}=1.0$, $g_{\rho
NN}^{T}=-1.2$,$g_{\rho\sigma\gamma}=-3.0$. In the last case we
consider $g_{\rho\sigma\gamma}$ as a fitting parameter, as well.

This allows to obtain some conclusions, concerning the proposed
model.

(1) First of all, we can see that
$|g_{\rho\sigma\gamma}|_{fit}\approx g_{\rho\sigma\gamma}$ of Ref.
\cite{R5}. Note that the sign of $g_{\rho\sigma\gamma}$ can not be
determined in the model of Ref. \cite{R5} because in that
model $d\sigma(\gamma
p\rightarrow p\rho^{0})/dt=|\pi|^{2}+|\sigma|^{2}$, i.e. there is no
$\pi\bigotimes\sigma$-interference contribution. But in the
model we consider it is possible in principle to determine
relative signs of coupling constants, for example, to the sign of
$\pi$-contribution, i.e. choosing for the product
$g_{\rho\pi\gamma}g_{\rho NN}>0$ the positive value.

(2) The resulting values for the coupling constants, $g_{\rho
NN}^{V}$ and $g_{\rho NN}^{T}$, are different from their standard
values,  which have been found early from NN-potential
\cite{R14,R15}, where typically, they are ranging from 2.97 to 3.16 for
$g_{\rho NN}^{V}$ and 12.5 to 20.8 for $g_{\rho NN}^{T}$ We must
repeat here once more that such difference could be considered as
natural due to the large difference in momentum transfer.

To characterize the quality of our fit, let us mention that the
$\chi^{2}$-value for the set of constants given above is $\chi^{2}/ndf=
2.1$ for the second set and 2.5 for the first set. We note that in
the interval $2.1 < \chi^{2}/ndf < 2.5$ there are a lot of
comparable minima in $\chi^{2}$ as a function of the three
coupling constants. This means a large correlation between these
constants, due to not so large sensitivity of $d\sigma/dt$ to the
details of the considered model.

Using both these fits we predict the t-dependence of
$d\sigma(\gamma p \rightarrow p \rho^{0})/dt$ for three another
energy intervals $1.19<E_{\gamma}<1.26$ GeV,
$1.26<E_{\gamma}<1.35$ GeV, $1.35<E_{\gamma}< 1.45$ GeV
which were not used in our fits, and we compare them with the
experimental results in Fig. 2 for two sets of $\rho NN$ and
$\rho\sigma\gamma$-coupling constants. In the same figure, we also show
our
fits for the other three energy intervals. One can see that both fits are
good enough
for all measured differential cross sections for $E_{\gamma}< 2$ GeV. Only
for the smallest energy with central value $E_{\gamma}=1.23$
GeV the predicted cross sections are larger than the experimental
values, such discrepancy could be considered as some indication of
possible contribution of nucleon resonances. The
contributions of different amplitudes to $d\sigma(\gamma p \rightarrow p
\rho^{0})/dt$ at
$E_{\gamma}=1.54$~GeV are shown in Fig. 3.

In any case, one can state that in the considered energy region
$E_{\gamma}< 2$ GeV the existing experimental data about
differential cross sections can be described relatively well in
the framework of a simple model with small number of adjustable
parameters. And the quality of the existing experimental data
allows different models without strong preference of some of them
as compared to many another possible models. Even our simplified
($\pi+\sigma+N$)model can be realized with at least by two
different possibilities with different values of $g_{\rho
NN}$ and $g_{\rho\sigma\gamma}$ coupling constants.

So, the available experimental data on $d\sigma(\gamma p
\rightarrow p \rho^{0})/dt$ is consistently well described in
($\pi+\sigma+N$)approach, and these data are not so discriminative
to the details of different possible such type models. Another
case is the polarization phenomena, even the simplest of them, for
example, the beam asymmetry $\Sigma$ induced by linear photon
polarization
\begin{equation}
 \Sigma=\frac{\sigma_{\parallel}-\sigma_{\perp}}{\sigma_{\parallel}
 +\sigma_{\perp}}
\end{equation}
with $\sigma_{\perp}(\sigma_{\parallel})$, induced by photon with
polarization orthogonal (parallel) to reaction plane, is sensitive
to reaction mechanism. Being equal to zero identically for
$(\pi+\sigma)$-exchange, this T-even polarization observable must be
discriminative to the relative value of other contributions. And
this is  demonstrated in Fig. 4 for proton target, where we present the
predicted values of $\Sigma$ for the SAPHIR energies. We see that
two sets of coupling constants result in $\Sigma$ which are different in
sign for the whole interval of t, demonstrating the importance
of the N-contribution, and especially $\sigma\bigotimes
N$-interference. So the future measurement of $\Sigma$-asymmetry
even at one energy and at one angle will be very decisive in the
choice of the correct reaction mechanism or in the set of
nonresonant background mechanisms.

Photoproduction of $\rho^{0}$-mesons on the neutron target,
$\gamma+n \rightarrow n+\rho^{0}$, will be interesting also,
especially in the near threshold region. Point here is that the
N-contribution, being controlled by the coupling constants,
$g_{\rho pp}^{V}$ and $g_{\rho pp}^{T}$, will be different due to
different electromagnetic characteristics of neutron and proton.
This also results in different interference $\sigma\bigotimes N$
contribution to all observables for processes, $\gamma+p
\rightarrow p+\rho^{0}$ and $\gamma+n \rightarrow n+\rho^{0}$,
such as the differential cross section and beam asymmetry.
The predicted behaviour of beam asymmetry for $\gamma+n \rightarrow
n+\rho^{0}$ is presented in Fig. 5.

We also note that the ratio of differential cross sections
($R=d\sigma(\gamma n \rightarrow n \rho^{0})/d\sigma (\gamma p
\rightarrow p \rho^{0})$ is sensitive to the set of vector
coupling constants, especially at large value of momentum transfer
$|t|$, with evident deviation of R from unity, in contrary directions
for the considered sets of coupling constants, as shown in Fig. 6.

Although we found good fit in our model, even two fits with
different sets of coupling constants, to the existing data about
differential cross sections for process, $\gamma+p \rightarrow p+
\rho^{0}$, in the near threshold region $E_{\gamma}< 2$ GeV, we
do not consider our results to be decisive. Indeed, we miss here
some "traditional" contributions, such as for example the
nucleon resonances in s-channel. There is no any strong
proof that the suggested model for the nonresonant
background is more suitable in comparison with another possible
approaches \cite{R1,R2,R3,R4,R5}. We can only say that our model is
relatively simple, being free from consideration of such
high-energy ingredients as the Pomeron exchange. In any case the
suggested model produces nontrivial polarization phenomena. We
note that the value of coupling constants $g_{\rho NN}^{T,V}$
which are controlled by N-contribution to the matrix element of
$\gamma+N \rightarrow N+\rho^{0}$ proces must be different
generally from the values for these constants that follow from
NN-interaction. And the values of $g_{\rho NN}$-coupling
constants in the time-like region of $\rho^{0}$-momentum can find a lot of
applications in consideration of
another process with vector meson production. Clearly, and not only for
our analysis, additional polarization data, about asymmetry
$\Sigma$ for example, will help to establish more uniquely the models
for $\gamma+N \rightarrow N +\rho^{0}$. And this is unavoidable as a
way in the solution of missing resonance problem.

\section{Conclusions}

Following are the main conclusions of our study and some general
remarks.

$\bullet$ It is shown that the relatively simple model
($\sigma+\pi+N$) can explain the SAPHIR data about the
differential cross section for $\gamma+p \rightarrow p+\rho^{0}$,
$E_{\gamma}< 2$ GeV, and therefore could be considered as a good
nonresonant background mechanism for searching the missing nucleon
resonances.

$\bullet$ Results for the VNN-coupling constants, their absolute
values and the signs as well, $g_{\rho
NN}^{T,V}$, considered in our work as a fitting parameters,
depend on the value of $g_{\rho\sigma\gamma}$-coupling constant.

$\bullet$ The simplest polarization observable, namely the beam
asymmetry $\Sigma$, is especially sensitive to possible variation
of parameters of our models in some limits, for which the
differential cross section is not so discriminative.

$\bullet$ The ratio of differential cross sections of
$\rho^{0}$-photoproduction on proton and neutron targets are
sensitive to the discussed variants of the suggested model.

In any case, the VNN-coupling constants from the fit to
$d\sigma(\gamma p \rightarrow p\rho^{0})/dt$ in the near threshold
region are different in absolute values and in signs from the so
called "standard" values of these constants, which have been
extracted from the data about NN-interaction or pion production,
$\gamma+N \rightarrow N+\pi$, due to essential difference in
momentum transfer.

In principle our estimation for the coupling constants $g_{\rho
NN}^{V}$ and $g_{\rho NN}^{T}$ will be useful for analysis of
another processes of vector meson production, such as for example,
$\pi+N\rightarrow N+V$, $N+N\rightarrow N+N+V$,
$e^{-}+N\rightarrow e^{-}+N+V$, $\overline{N}+N\rightarrow \pi+V$,
$\overline{N}+N\rightarrow \gamma+V$, $\overline{N}+N\rightarrow
V+V$ etc., in the framework of the Effective Lagrangian approach.

$\bullet$ Our fitting procedure demonstrates that even the
differential cross sections are sensitive to the relative sign of
the different contributions to the matrix element of the process,
$\gamma+p\rightarrow p+\rho^{0}$.

$\bullet$ Cut-off parameters $\Lambda_{i}$ of above mentioned
phenomenological form factors, which typically must be introduced
in electromagnetic and strong vertexes of the considered pole
diagrams, must be object of special consideration, because such
universality i.e., applicability of the same form factors for
different process is not proved rigorously. It is only a very
simplified procedure.

\section*{Acknowledgement}

We thank M. P. Rekalo for suggesting this problem to us and we gratefully
acknowledge his guidance and fruitful discussions during the course
of our work.

\newpage
\begin{figure}
$\left. \right.$
\vskip 15cm
    \includegraphics{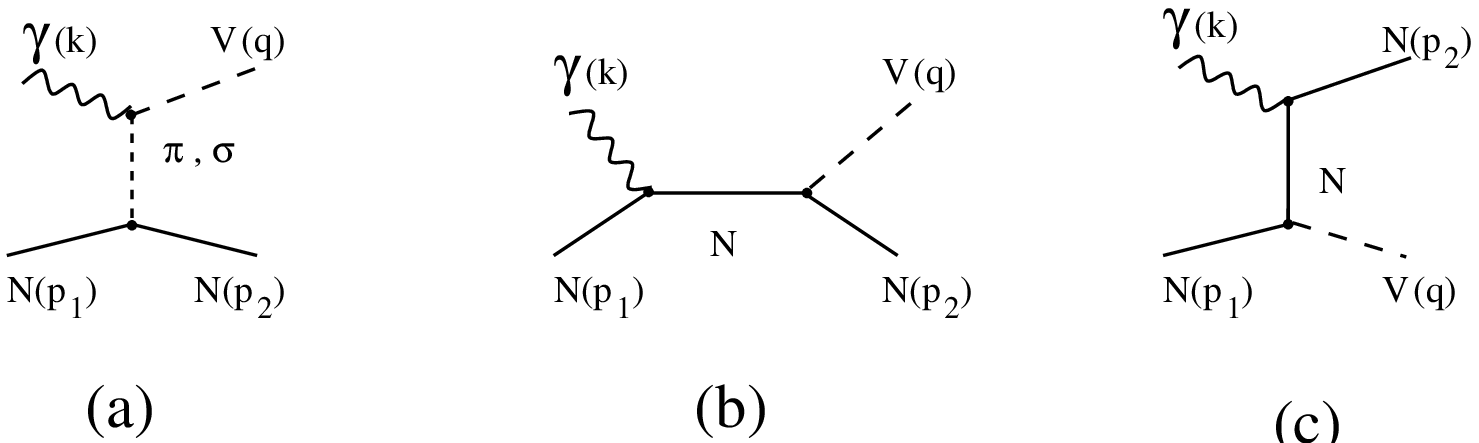} \vskip -1cm \caption{Mechanisms of
the model for $\rho^{0}$-photoproduction: (a) t-channel exchanges,
(b) and (c) s- and u-channel nucleon exchanges.}
\end{figure}
\newpage
\begin{figure}
$\left. \right.$
\vskip 10cm
   \includegraphics{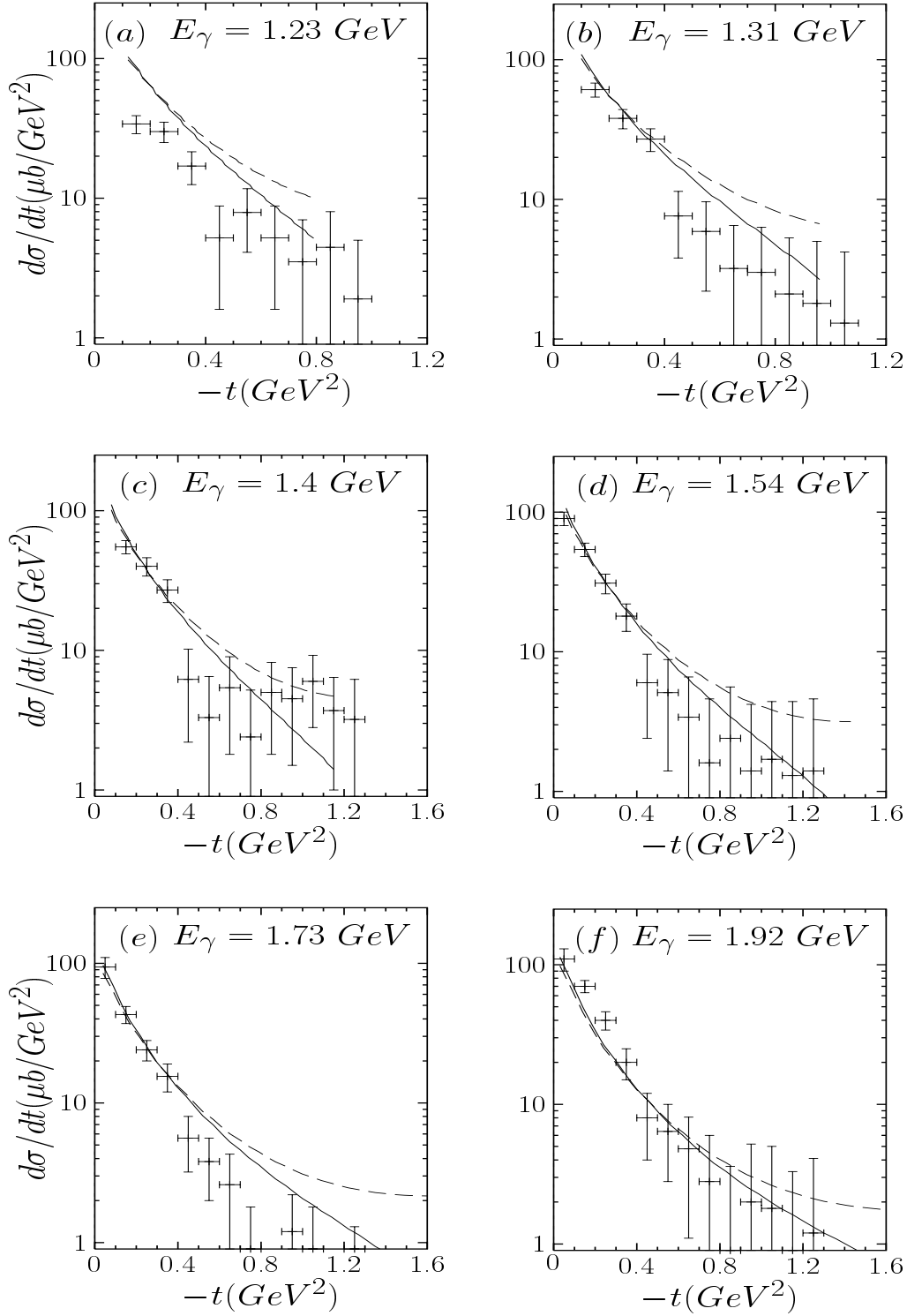} \vskip 9.5cm \caption{Comparison of
experimental differential cross section  data for $\gamma +p \rightarrow
p+\rho^{0}$
at $E_{\gamma}=1.23$, $1.31$, $1.4$, $1.54$, $1.73$ and $1.92$ GeV
from [11] with the calculation of suggested model. Solid and
dashed lines correspond to $g_{\rho NN}^{V}= 1.0$, $g_{\rho
NN}^{T} =-1.2$, $g_{\rho\sigma\gamma}=-3.0 $, and $g_{\rho
NN}^{V}=0.4$, $g_{\rho NN}^{T}=1.0$, $g_{\rho\sigma\gamma}=2.7$,
respectively.}
\end{figure}
\newpage
\begin{figure}
$\left. \right.$
\vskip 10cm
    \includegraphics{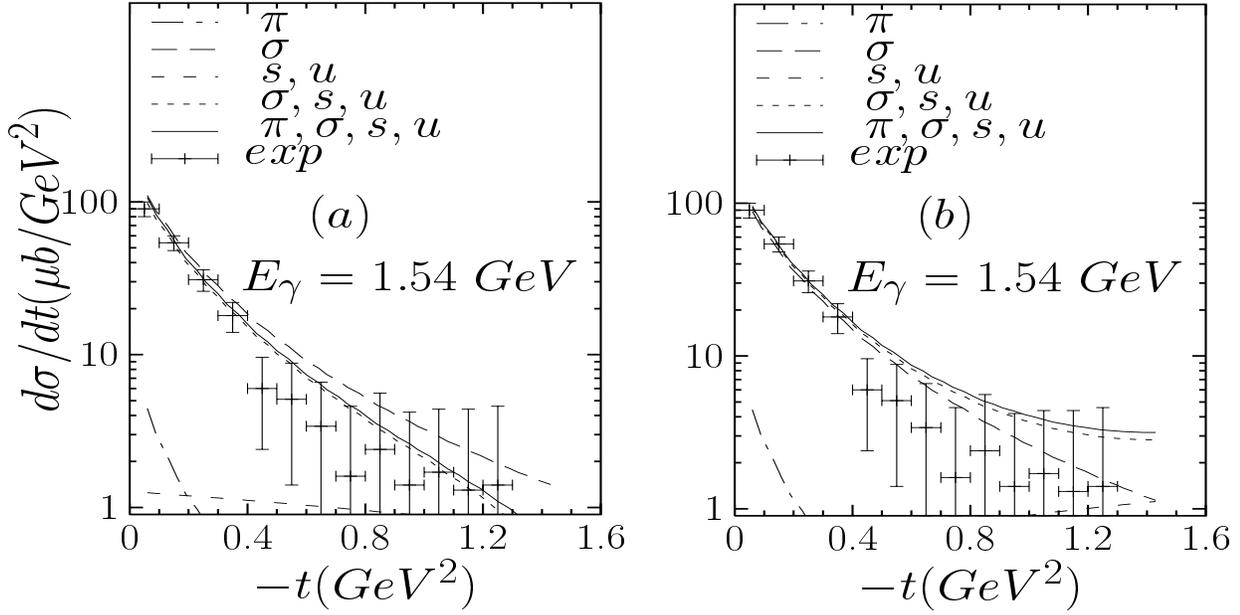}
\vskip 5.5cm
\caption{ Different contributions to the differential
cross sections of $\gamma+p\rightarrow p+\rho^{0} $ at
$E_{\gamma}=1.54$ GeV  for two different fitted parameter
values:(a) $g_{\rho NN}^{V}=1.0,~g_{\rho
NN}^{T}=-1.2,~g_{\rho\sigma\gamma}=-3.0 $~~(b) $g_{\rho
NN}^{V}=0.4,~g_{\rho NN}^{T}=1.0,~g_{\rho\sigma\gamma}=2.7 $.}
\end{figure}
\newpage
\begin{figure}
$\left. \right.$
\vskip 10cm
    \includegraphics{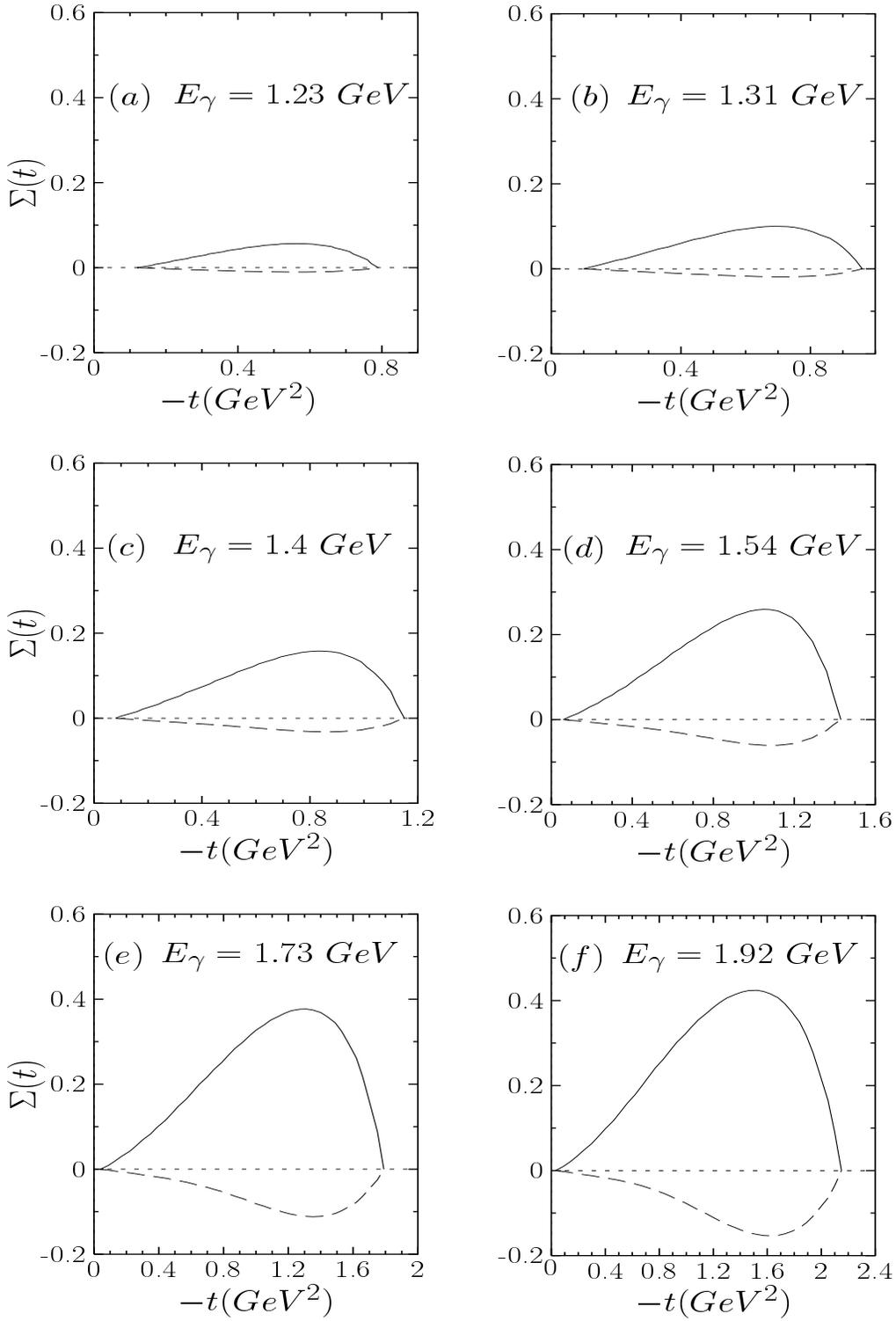}
\vskip 9.5cm
\caption{ Predicted behaviour of beam asymmetry for
$\gamma+p\rightarrow p+\rho^{0}$ at $E_{\gamma}=$ $1.23$, $1.31$,
$1.4$, $1.54$, $1.73$ and $1.92$~GeV. Solid and dashed lines
correspond to $g_{\rho NN}^{V}=1.0,~g_{\rho
NN}^{T}=-1.2,~g_{\rho\sigma\gamma}=-3.0 $, and $g_{\rho
NN}^{V}=0.4,~g _{\rho NN}^{T}=1.0,~g_{\rho\sigma\gamma}=2.7 $,
respectively.}
\end{figure}
\newpage
\begin{figure}
$\left. \right.$
\vskip 10cm
    \includegraphics{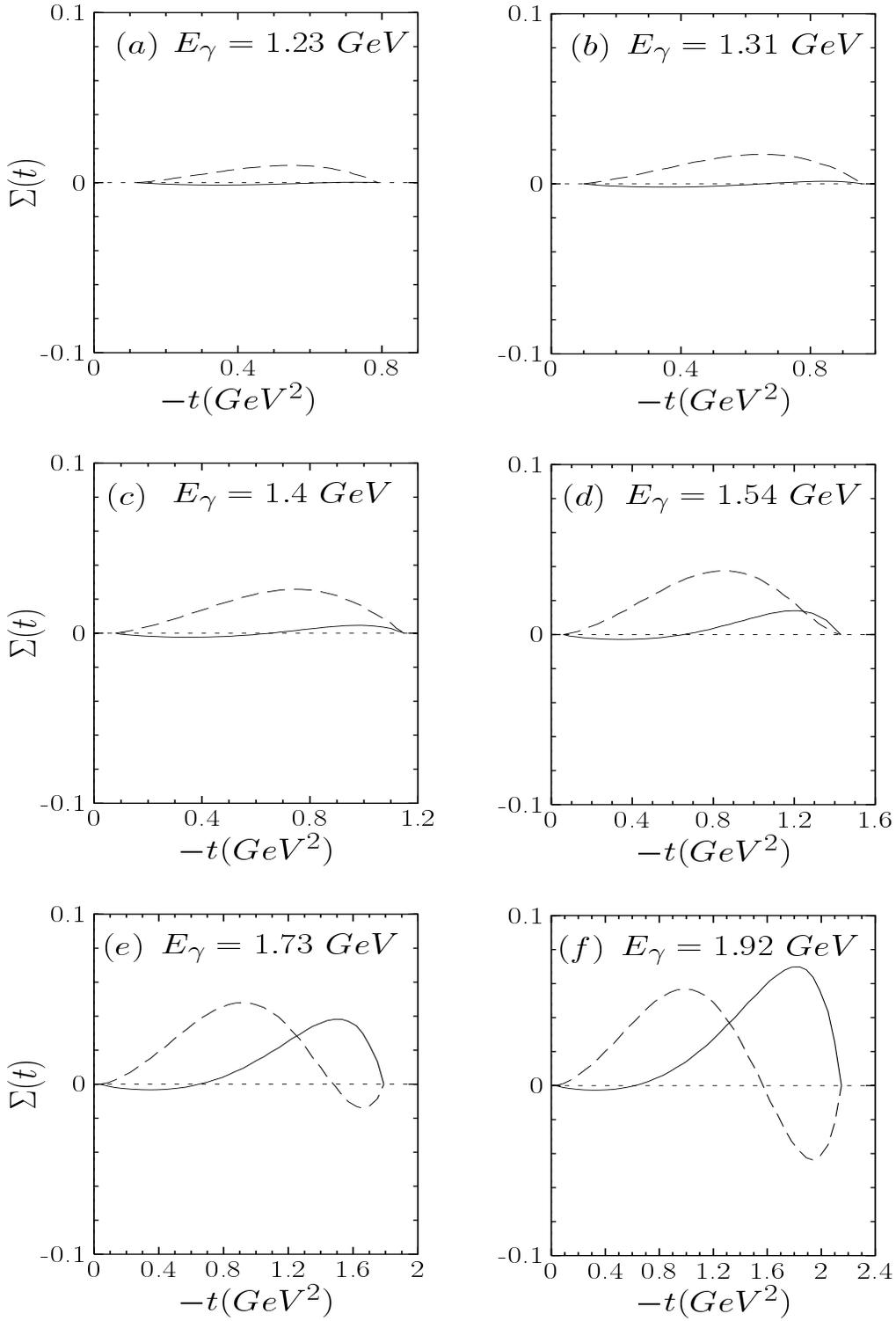}
\vskip 9.5cm
\caption{Predicted behaviour of beam asymmetry for $\gamma
+n\rightarrow n+\rho^{0}$ at $E_{\gamma}=$ $1.23$, $1.31$, $1.4$,
$1.54$, $1.73$ and $1.92$~GeV. Notation for different graphs is the same
as in
Fig.4.}
\end{figure}
\newpage
\begin{figure}
$\left. \right.$
\vskip 10cm
    \includegraphics{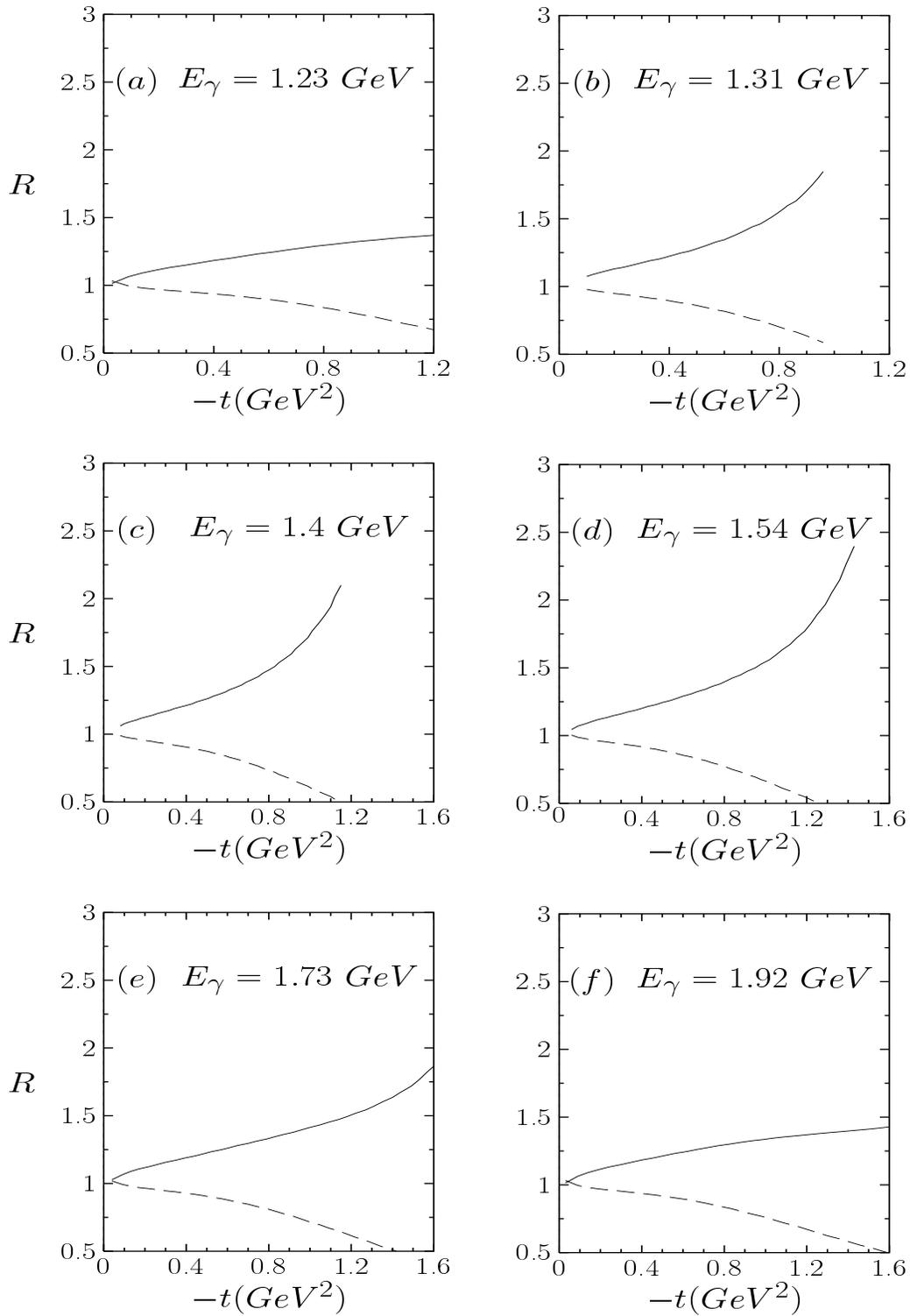}
\vskip 9.5cm
\caption{Ratio of differential cross section on neutron and
proton target ($R=$ $d\sigma(\gamma n\rightarrow n
\rho^{0})/$ $d\sigma(\gamma p\rightarrow p \rho^{0})$ ) at
$E_{\gamma}=$ $1.23$, $1.31$, $1.4$, $1.54$, $1.73$ and $1.92$~GeV
with the total contributions of exchange mechanisms ($\pi$,
$\sigma$, s, u). Notation for different graphs is the same as in Fig.4.}
\end{figure}

\end{document}